# A COMPARISON BETWEEN TWO SIMPLE MODELS OF A SLUG FLOW IN A LONG FLEXIBLE MARINE RISER

*A. Pollio*[1], *M. Mossa*[2]

[1] Corresponding author, PhD, grant recipient, Technical University of Bari, Department of Water Engineering and Chemistry, Via E. Orabona 4, 70125 Bari, Italy, e-mail: a.pollio@poliba.it, fax: +39 080 596 3414.

[2] Professor, PhD, IAHR member, ASCE member, Technical University of Bari, Department of Environmental Engineering and Sustainable Development, Via E. Orabona 4, 70125 Bari, Italy, e-mail: m.mossa@poliba.it, fax: +39 080 596 3414.

## ABSTRACT

Slug flows are extremely interesting multiphase regime phenomena which frequently occur in flexible marine risers used by the petroleum industry in offshore environments and have both a liquid and gaseous phase. Generally, the gaseous phase is pumped in together with the liquid phase to facilitate the suction of the latter. Literature and experimental observations show a wide range of multiphase regimes which are the result of numerous parameters such as liquid and gaseous discharge, the pipe diameter and its inclination.

This paper describes two simple models of the slug flow regime by means of an equivalent monophase flow with a non-constant density.

The slug regime is modelled as a monophase density-varying flow with a sinusoidal density, travelling along the pipe itself towards the top end node of the riser. Starting from the bottom end, it is characterized by adiabatic processes and energy loss along the entire length of the pipe. In the first model, the slug wavelength is supposed to be independent of the riser inclination, while in the second one a simple linear relationship between the slug wavelength and the pipe inclination was imposed.

The global equation of the motion of the riser (written in a two-dimensional domain throughout the plane containing the riser) was solved using a Matlab code in the time domain, starting with a free-hanging elastic catenary as the initial configuration to investigate the dynamic stresses which the riser undergoes. In particular, the axial tensile force, the bending moment, the viscous structural damping, the wave-induced forces and the riser-seabed interaction are all modelled here.

This work presents a comparison between the two models in two main kinds of configuration (a very long riser with and without seabed presence) and it allows the authors to make some considerations on general pipe behaviour.

## INTRODUCTION

Recently, the oil industry has shown interest in extracting coal oil from oilfields located at deep ocean levels. To achieve this goal, flexible marine risers, which are specifically-designed flexible pipes, must be used in the extraction operation. The risers allow for the transfer of crude oil from the seabed, where there is a pumping station, up to the free water surface where other structures (modified oil tanks or oil platforms) are located in order to store the crude oil before treatment. Because of the extreme depths involved (generally from several hundred up to thousands of meters), the risers are subjected to numerous forces that induce high stress levels in the pipe material. Therefore, an appropriate pipe design is essential to render them more resistant and to avoid possible breaks and subsequent dispersal of oil and gas which could be transported towards sensitive areas near coastal environments.

Two examples of forces acting upon marine risers are the external forces caused by wave motion and sea currents and the inner forces due mainly to a unique multiphase flow known as the slug flow regime. The inner flow into the riser consists of a liquid phase (crude oil) and of a gaseous phase (used to increase the rising flow velocity in the pipe). However, other fundamental boundary forces must be taken into account to analyze riser behaviour in the environment in which it works. The risers are also subjected to the motion of the platform to which it is joined at the water surface, the effects of seabed contact and the action of the buoys (in this case it must be highlighted that buoys distributed along the riser are used in order to pull the tube away from the seabed and to decrease the stress levels within the riser).

In this study, in order to investigate riser behaviour in the time domain, the lumped mass approach is used to represent the riser and all the forces acting on it. This technique allows for a very simple formulation of the equation of the motion of the cable. Furthermore, it allows for good results in comparison with the other approaches available in literature and it is also much less expensive to compute. Specifically, the riser is divided into a given number of segments and nodes and at each node a lumped mass representing the segment mass is associated. All the forces acting on the pipe are applied at the various nodes by means of the free body diagram of each lumped mass. In particular, in this study the modelled forces are the submerged weight of the riser, the Morison's forces in the presence of regular monochromatic waves, the slug flow forces and the seabed-riser interaction. The inner stresses of the pipe are added to these: the axial tensile force, the tangential structural damping modelled by a viscous model and the shear force, useful in investigating the bending moment distribution of the riser. The presence of buoys distributed along the riser are not considered in this study; neither is the free surface platform motion.

In this paper, the authors present some results from the time domain simulations for two riser configurations (either in the absence or presence of the seabed) for which both its ends are pinned as a free hanging configuration. In order to obtain these results, the non-linear equation of the motion was solved by using the ODE 113 algorithm implemented in MATLAB.

The results are presented in order to highlight the impact of the inner hydraulic regime due to the slug flow on the internal stresses of the riser. Several considerations could be taken into account by the designers of risers in planning a suitable riser resistance, thus avoiding possible breaks.

## STATE OF THE ART

Researchers have carried out a great deal of work on the study of cable behaviour. Some of these studies also concern the analysis of marine pipes subjected to external forces either in deterministic or random fields, yet only a few have focused on marine flexible risers considering the effects of internal flows, and in particular intermittent multiphase flows. Multiphase flow regimes are a very complex area of study. A general theoretical approach leads to a series of complicated equations that also takes heat transfer into account. In the context of a marine riser, it can be extremely difficult to describe these equations as they require numerous parameters. Furthermore, if the objective is to model the slug flow pattern, it must be described in a specific way as it is an intermittent and unsteady phenomenon. Generally, the equations describing this non-steady motion are highly complex and as a result some simplifications may be useful to describe simple models, an approach adopted in this study.

The time domain analysis is the most accurate method available to study the behaviour of a flexible marine riser. This is because only this type of analysis is capable of describing all the non-linearities in structural geometry, loading and material behaviour in a suitable way. Although it requires more computer time, it should potentially provide greater accuracy than its frequency domain equivalent. Typical methods used by researchers include the finite difference, and the Wilson and Newmark methods. Many people in offshore applications use Newmark's method to integrate the governing equations in time due to its stability, accuracy and easiness of implementation.

Gardner and Kotch (1976) used a finite element analysis in conjunction with the Newmark method to provide a time domain technique for the analysis of vertical risers in waves. Equations of dynamic equilibrium in time have been developed using a stiffness proportional damping matrix. They supposed that the effects of structural damping in comparison with fluid damping are slight and that it is more important to stabilize the problem numerically.

It should be noted that in the code used in our study, the introduction of structural damping allowed the authors to filter the output of the various quantities involved in the system in order to obtain the same results as the commercial code Orcaflex. The difference in our approach is that the coarse code (that is, without structural damping) does not create numerical difficulties but takes into account the higher frequency content; as a result, the outputs are noised.

Cowan and Andris (1977) also used the Newmark method for a time domain analysis of a pipe laying system. In the first step (static analysis) they used a modified form of Newton's method followed by a time domain analysis which considered the dynamic variations of the system parameters as perturbations of their mean static values.

Natvig (1980) used the Wilson method to introduce artificial damping of the higher modes of vibrations allowing larger time steps to be used for integration.

Patel and Jesudasen (1987) considered vertical free hanging risers subjected to vessel motion and wave loading using the Newmark method. Firstly, they used an initial incremental static analysis performed with the stiffness matrix updated at each stage. They then they performed an eigenvalue analysis using a diagonal mass matrix.

Patel and Sarohia (1984) used frequency and time domain procedures for vertical risers. The frequency domain solution employed a linearized drag force and Rayleigh structural damping was assumed in the first two modes of vibrations. The time domain solutions employed the Newmark method. Most researchers are reluctant to become involved in the detailed mathematics of more advanced methods even in cases where these have been shown to save on both computer effort and storage. Although the choice of the method affects the ease of programming, the accuracy of the results for a given time step and the run time may be improved by reducing the time step.

A clear presentation of the derivation of the equation of the motion of a riser by using a lumped mass approach may be found in Ghadimi (1988) who implemented the Newmark technique to solve the problem. In his work, he used the tangent stiffness matrix and also a model for the bed force by a hyperbolic tangent function to avoid numerical instabilities, but the inner effects of a slug flow were not considered.

The most interesting research on the internal flow effect in a riser by considering the actual features of the phenomenon was carried out by Patel and Seyed (1989). In their work, the authors derived the equation by following the finite element approach and the slug flow was modelled by considering a fluid with a periodic density. Such a model is also implemented in this study, but it is used by considering the different lumped mass technique solved with an ODE algorithm and also adding a viscous structural damping model; this will be more fully explained below.

More literature about the study of the equation of the motion for flexible marine risers may be found in Patel (1995) in which all the time and frequency techniques are clearly described together with all their advantages and disadvantages. The good quality of the code presented in this work with regard to the comparison between its results and those obtained from the commercial code Orcaflex is discussed in Pollio et al. (2006).

## DESCRIPTION OF THE PRESENT MODEL

In this section the riser model used in this study is explained, together with its equation of motion carried out by following a lumped mass technique in which the riser is compounded by a set of nodes connected to each other by massless elements. Considering each node as isolated, the free body diagram allows for the forces acting on it, as shown in Figure 1. Essentially, these forces are of two types: the forces spreading throughout the pipe wall, that is, the shear force $\mathbf{Q}_i$, the structural damping $\mathbf{F}_i^c$ and the axial force $\mathbf{T}i$, and the other forces, due to external causes (weight $\mathbf{W}_i$, Morison's effects $\mathbf{F}_i^D + \mathbf{F}_i^I$) and the internal force (slug effect $\mathbf{F}_i^S$). We will derive the latter force from the global equation of the dynamic equilibrium.

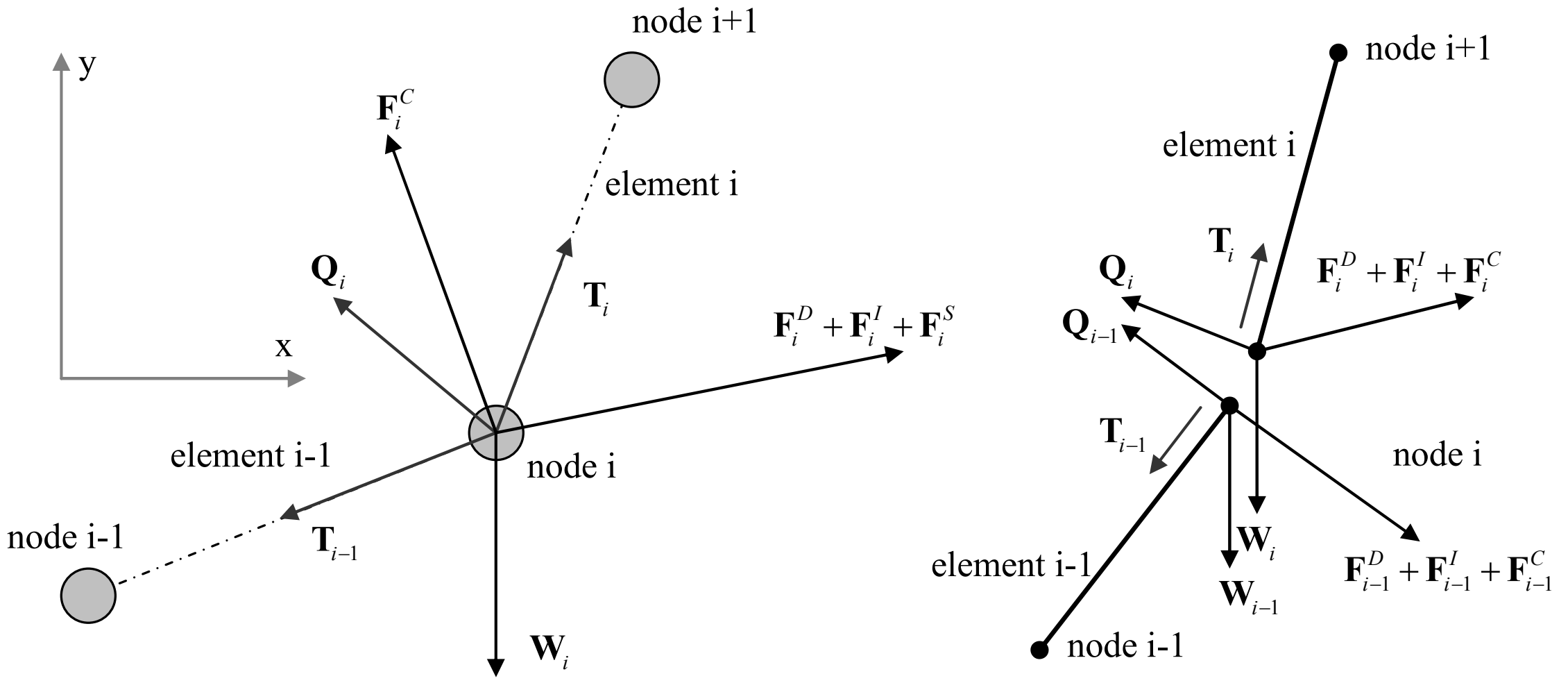

Figure 1. Vectorial representation of the forces acting on an element riser.

Since the riser model is bidimensional, the node positions are represented by their horizontal and vertical coordinates $x_i$, $y_i$ in a global reference system.

The total force acting at element $i$ is given by the sum of the forces acting on nodes *i-1* and $i$ belonging to the same element:

$$\mathbf{F}_i = \mathbf{T}_{i-1} + \mathbf{T}_i + \mathbf{W}_i + \mathbf{F}_i^D + \mathbf{F}_i^I + \mathbf{F}_i^S + \mathbf{F}_i^C + \mathbf{Q}_i \qquad (1)$$

Since each node belongs to two riser element, the authors preferred to write an equation of the motion to each element, and in doing this, for a single element it is obtained a vectorial equation of the motion having the dimension $4\times 1$ is obtained, describing the force components along the $x$ and $y$ axis at the extreme nodes of the element. The subsequent step is to assemble the single equation of motion for the elements across the whole system.

Generally, if node $i$ belongs to element $i$, the force equilibrium becomes:

$$\mathbf{F}_i = \mathbf{T}_i + \mathbf{W}_i + \mathbf{F}_i^D + \mathbf{F}_i^I + \mathbf{F}_i^C + \mathbf{Q}_i \qquad (2)$$

in which the slug force does not appear as it will be added in a subsequent step.

The first side term of equation (2) may be expressed as the product between the mass and the acceleration at the node and it must take into account the structural mass and the so-called added mass due to the marine water environment:

$$\mathbf{F}_i = m^s \ddot{\mathbf{S}}_i + \rho_w A(c_{in} - 1)\left(\ddot{\mathbf{S}}_i\right)_n \tag{3}$$

where $\mathbf{S}_i$ is a vector of the node coordinates and the double dots indicate its derivative with respect to time, $c_{in}$ the inertia coefficient, while $A$ is the pipe transversal area (see equation (8)); note that only the normal component of the added mass term was considered because the tangential one is supposed to be negligible. For node $i$ belonging to element $i$, the normal component of the acceleration is:

$$\left(\ddot{\mathbf{S}}_i\right)_n = \ddot{\mathbf{S}}_i - \left(\mathbf{t}_i^T \cdot \ddot{\mathbf{S}}_i\right)\mathbf{t}_i = \ddot{\mathbf{S}}_i - \left(\mathbf{t}_i \cdot \mathbf{t}_i^T\right)\ddot{\mathbf{S}}_i \tag{4}$$

and then, after rearranging:

$$\mathbf{F}_i = m^s [\mathbf{I}]\ddot{\mathbf{S}}_i + m_i^a [\mathbf{B}]_i \ddot{\mathbf{S}}_i = [\mathbf{M}]_i \ddot{\mathbf{S}}_i \tag{5}$$

where [**I**] is the identity matrix $2\times 2$ and:

$$m_i^a = \rho_w A(c_{in} - 1) m^a \frac{L_i}{2} \tag{6}$$

in which $m^a$ is the normal added mass per unit length, $A$ is the area cross-sectional to the inertial force, $\rho_w$ the seawater density, $L_i$ the element length and

$$[\mathbf{B}]_i = \begin{pmatrix} 1 - t_{i,x}^2 & -t_{i,x}\, t_{i,y} \\ -t_{i,x}\, t_{i,y} & 1 - t_{i,y}^2 \end{pmatrix} \tag{7}$$

is a unit vector matrix with information about the orientation of the element (note that the subscripts indicate the element number and the direction of the vector component respectively).

If $D_i$ and $D_e$ indicate the respective inner and outer diameters, then:

$$A = \frac{\pi}{4}\left(D_e^2 - D_i^2\right) \tag{8}$$

Thus the equation of motion for node *i* belonging to element *i* is:

$$\left[\mathbf{M}\right]_i \ddot{\mathbf{S}}_i = \mathbf{T}_i + \mathbf{W}_i + \mathbf{F}_i^D + \mathbf{F}_i^I + \mathbf{F}_i^C + \mathbf{Q}_i \tag{9}$$

In the same way, if node *i* is seen as belonging to element *i-1*, a similar equation can be written.

The module of the axial force is expressed as:

$$T_j = E\,A\frac{L_j - L_{j0}}{L_{j0}} \tag{10}$$

where *EA* is the riser axial stiffness, $L_j$ is the instantaneous element length and $L_{j0}$ the initial length of the same element. In all the tests carried out in this study, the so-called *effective tension* is shown. For a tube submerged in fluid with pressure pext and containing fluid with pressure *p*, density $\rho_o$ and moving with a mean velocity *U*, the effective tension $T_{eff}$ is defined by the expression:

$$T_{eff} = T_{wall} + p_{ext}\,A_e - p\,A_i - \rho_o\,A_i\,U^2 \tag{11}$$

where $A_e$ is the pipe outer area and $A_i$ its inner area.

The effective tension must be taken into account when studying a riser placed in fluid with pressure varying along its depth (as in the case of seawater). It shows that together with the real tension $T_{wall}$ acting on the pipe, there is also  external and internal hydrostatic pressure and dynamic force due to the internal fluid. The equation also demonstrates that external pressure acts as a compressive axial force because in order to maintain constant effective pressure, any increase in internal pressure requires an external pressure increase in the absence of inner fluid velocity, which still shows an inner fluid configuration. It is important to note that the velocity of the fluid in the pipe affects the wall tension and that the magnitude of the variation of the wall tension is independent of the flow direction, because there is a square mean velocity. In the Matlab code presented in this study, the actual tension is that calculated at the end of each temporal step because it is obtained by considering all the effects resulting from outer environmental pressure, inner fluid pressure and the fluid velocity in the pipe.

$W_i$ is the submerged weight of the riser elements $i$ lumped at the considered node $i$, where its horizontal component is zero and the vertical one is expressed as:

$$w_i = \left( m^s - \rho_w \frac{\pi D_e^2}{4} \right) g \frac{L_i}{2} \tag{12}$$

in which the buoyancy is considered, $\rho_w$ represents the seawater density, $g$ gravity acceleration and $m_s$ is the structural mass per unit length of the riser.

A general form of the Morison equation may be found in Patel and Sarohia (1984) representing the hydrodynamic force per unit length acting on a slender member in presence of waves and steady current. In this work a three-dimensional form of the equation is considered (Brebbia and Walker, 1979). The Morison force acting upon a pipe may generally be written as the sum of two terms, here shown as $\mathbf{F}_i^I$ and $\mathbf{F}_i^D$ .

$\mathbf{F}_i^I$ is the induced inertia force as proposed by Morison et al. (1950) which represents the inertia component deriving from wave and current phenomena and considers the mass acceleration and the added mass terms. $\mathbf{F}_i^D$ is the induced drag force which depends on the relative velocity between the riser element and the water particle velocity together with the marine current velocity.

The Morison forces can be referred to a local reference system for the element, taking into account the normal, binormal and tangential directions.

In this case there is no third dimension, thus the binormal component is not considered.

In this study, the tangential component of the drag force will not be considered because the tangential Morison drag coefficient is smaller than the normal ones (from 30 to 120 times less (as shown by Brebbia and Walker, 1979)): physically, this is due to the fact that the tangential relative motion between the wave-current and the riser element induces a neglecting friction.

Furthermore, the water particle velocity is expressed following Airy's linear theory for deep water. In this case, to avoid numerical instabilities, the following approximation is used for the water particle velocities along the $x$ and $y$ axis:

$$\begin{cases} u_x = \frac{H}{2}\frac{gT}{L}\exp\left(\frac{2\pi}{L}y\right)\cos\left(\frac{2\pi}{L}x - \frac{2\pi}{T}t\right) \\ u_y = \frac{H}{2}\frac{gT}{L}\exp\left(\frac{2\pi}{L}y\right)\sin\left(\frac{2\pi}{L}x - \frac{2\pi}{T}t\right) \end{cases} \tag{13}$$

where $u_x$ and $u_y$ are the $x$ and $y$ components of the water particle velocity, $H$ is the wave height, $T$ is the wave period, $L$ is the wavelength and $t$ the time.

The shear force $\mathbf{Q}_i$, connected to the bending momentum, is expressed by following the Euler-Bernoulli beam theory and it is assumed to be constant along each element. The approximate bending moment at the generic node is obtained by using the Euler-Bernoulli beam theory (see also Ghadimi, 1988):

$$\mathbf{M}_m = E\,J\,R_m\,\mathbf{b}_m \tag{14}$$

in which the subscript $m$ indicates the generic node $m$, as shown in Figure 2,

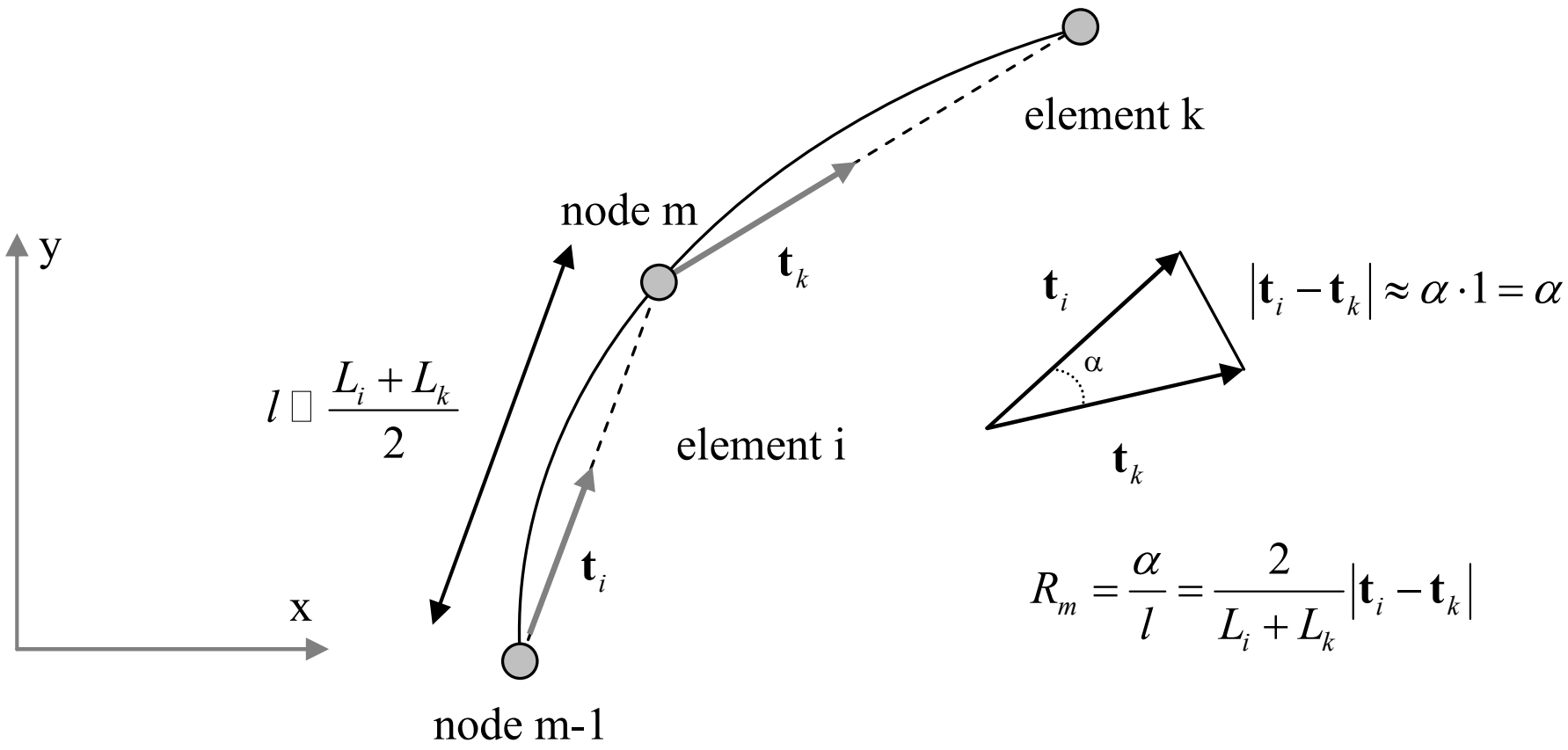

Figure 2. Evaluation of the discrete curvature at the generic node $m$.

and where:

$$R_m = 2\frac{|\mathbf{t}_k - \mathbf{t}_i|}{L_k + L_i} = \frac{2}{L_k + L_i}\left[ \left( t_{k,x} - t_{i,x} \right)^2 + \left( t_{k,y} - t_{i,y} \right)^2 \right]^{1/2} \tag{15}$$

$$\mathbf{b}_m = \mathbf{t}_{ki} \times \mathbf{n} \tag{16}$$

$$\mathbf{t}_{ki} = \frac{L_k\,\mathbf{t}_i + L_i\,\mathbf{t}_k}{L_k + L_i} = \frac{1}{L_k + L_i}\left[ L_k \left( t_{i,x}, t_{i,y} \right) + L_i \left( t_{k,x}, t_{k,y} \right) \right] = \frac{1}{L_k + L_i}\left( L_k\, t_{i,x} + L_i\, t_{k,x}, L_k\, t_{i,y} + L_i\, t_{k,y} \right) \tag{17}$$

$$\mathbf{n} = \frac{\mathbf{t}_k - \mathbf{t}_i}{|\mathbf{t}_k - \mathbf{t}_i|} = \frac{\left( t_{k,x} - t_{i,x}, t_{k,y} - t_{i,y} \right)}{\left[ \left( t_{k,x} - t_{i,x} \right)^2 + \left( t_{k,y} - t_{i,y} \right)^2 \right]^{1/2}} \tag{18}$$

$R_m$ is an estimate of the pipe curvature at point *m*, whilst $b_m$ is the binormal unit vector at the same point, defining the direction of the discrete momentum. The shear force on the *i-th* element depends on the difference between the moments at the extreme nodes *m-1* and *m* as calculated by equation (14):

$$\mathbf{Q}_i = \mathbf{t}_i \times \frac{\left( \mathbf{M}_m - \mathbf{M}_{m-1} \right)}{L_i} \tag{19}$$

To obtain the structural damping force $\mathbf{F}_i^c$ for a two dimensional system such as that under consideration, let us consider a set of *k* masses connected by axial springs and tangential viscous damping effects.

If $(c_i,k_i)$ represents the axial viscous damping and stiffness of each element of the riser and $\xi_i$ represents the critical damping ratio (Raman-Nair and Baddour, 2003) , then:

$$c_i = 2\,\xi_i \sqrt{k_i\, m_i} \tag{20}$$

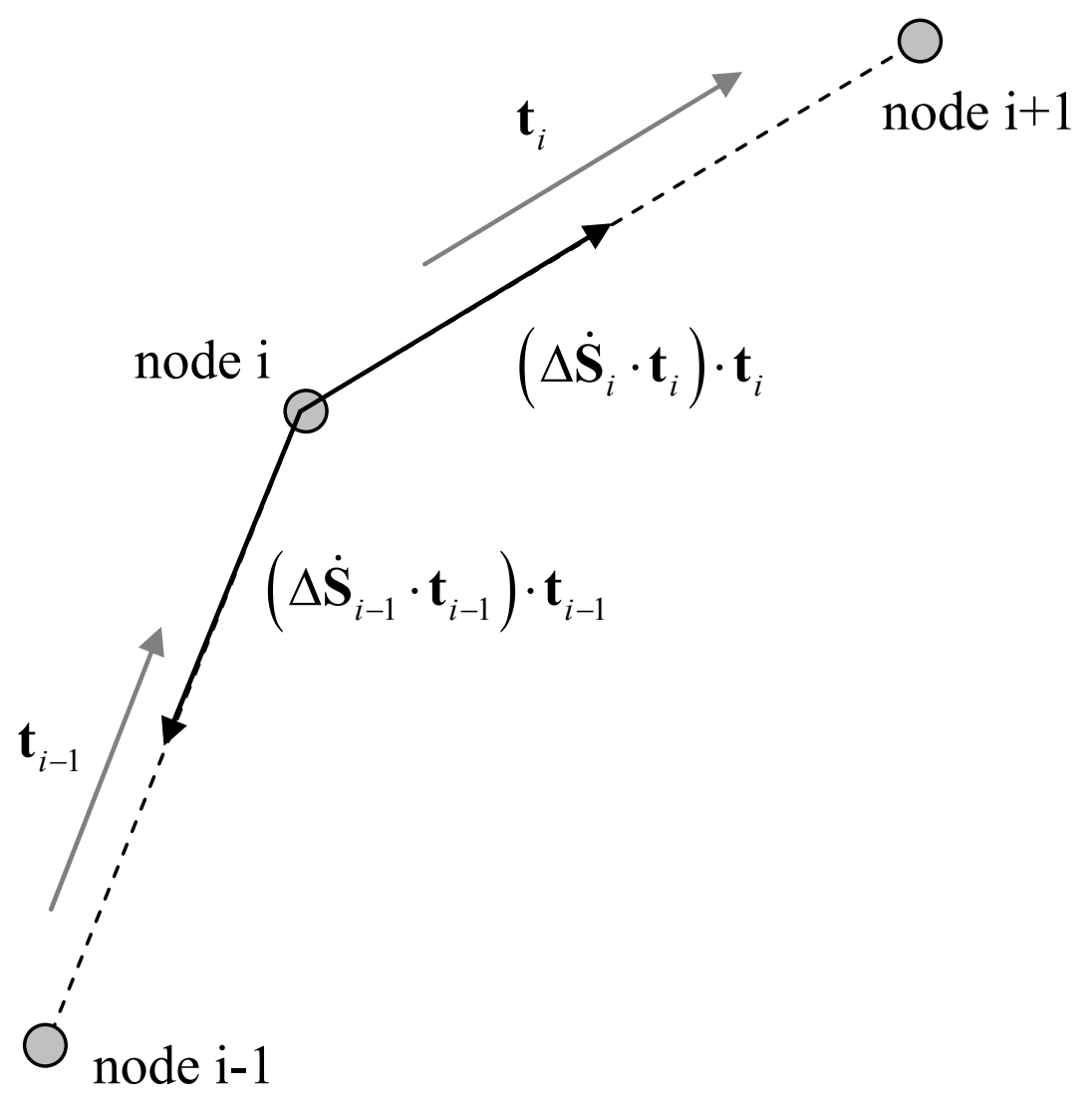

Figure 3. Modelling of the structural damping

Figure 3 shows the structural damping forces acting along a generic element. In a free body diagram these forces may be seen applied at each node and they act along the element in the opposite direction to the relative displacement rate of the two contiguous nodes. If $S_i=(x_i,y_i)$ is the displacement vector of each node of the *i-th* element, the force due to the structural damping on the node $i$ may be expressed, in terms of components, as:

$$\mathbf{F}_i^c = -c_{i-1}\left(\Delta\dot{\mathbf{S}}_{i-1}\cdot\mathbf{t}_{i-1}\right)\mathbf{t}_{i-1} - c_i\left(\Delta\dot{\mathbf{S}}_i\cdot\mathbf{t}_i\right)\mathbf{t}_i \tag{21}$$

In particular, the product $\left(\Delta\dot{\mathbf{S}}_i\cdot\mathbf{t}_i\right)\mathbf{t}_i$ represents the relative velocity between the nodes $i$ and $i+1$ along the direction of the element $i$. For this study, the critical damping factor $\xi_i$ was always assumed as 0.2 after a proper calibration with the Orcaflex commercial code.

Equation (9) (in which the slug force does not yet appear), written for both nodes, can be assembled in the following vectorial equation ($4\times1$) related to the *j-th* element:

$$\left[\mathbf{M}\right]_j \ddot{\mathbf{S}}_j = T_j\,\boldsymbol{\upsilon}_j + \mathbf{W}_j + \mathbf{F}_j^D + \mathbf{F}_j^I + \mathbf{F}_j^C + \mathbf{Q}_j \tag{22}$$

where υ denotes a $4\times 1$ vector containing the unit vector components of element *j* (see also Ghadimi, 1988).

The seabed acts only on those riser nodes actually lying on it. In this study, two components of the force deriving from the contact between the seabed and the riser were considered: the first, $\mathbf{R}^s$, is due to the elastic behaviour of the seabed which affects the nodes lying on it while the second, $\mathbf{F}^s$, is a result of friction and acts along the seabed surface (see Figure 4).

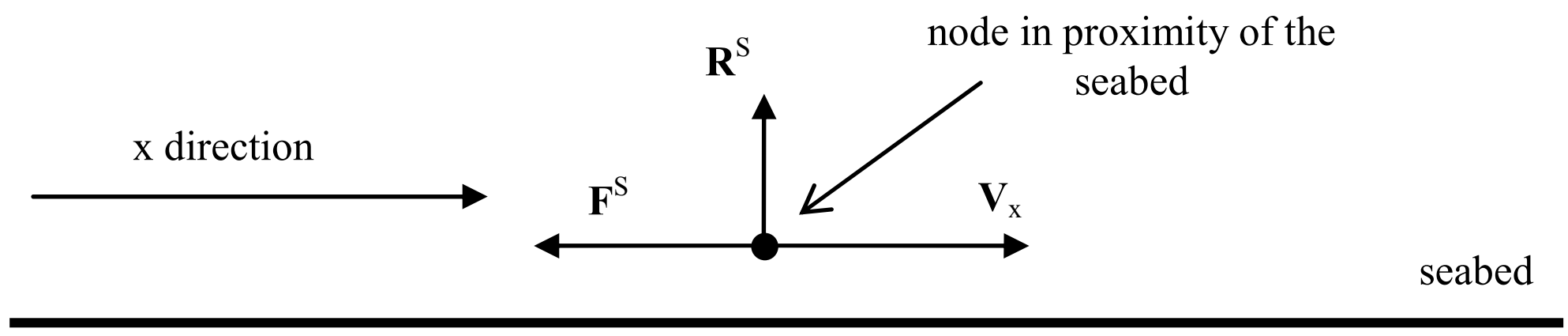

Figure 4. The seabed-riser interaction is seen as the sum of an upward reaction force $\mathbf{R}^\mathbf{s}$ and a horizontal friction force $\mathbf{F}^\mathbf{s}$.

The normal reaction $\mathbf{R}^s$ is written as function of the relative displacement between the seabed (with depth *d*) and the node depth *y* and also depends on the seabed stiffness $e=E_sA_s/d \propto W$ (represented by the stiffness factor *e* which here is considered proportional to the submerged weight of the pipe) where $E_s$ is Young's module of the soil and $A_s$ is the contact area between the pipe and the seabed:

$$\mathbf{R}^s = e\frac{(y-d)}{d}\mathbf{k} \tag{23}$$

where **k** is the unit vector pointing upwards.

It is clear that the normal reaction of the bed is zero when $y=d$. It reaches its maximum value as the node is placed under the bottom and acts in an upward direction opposite to the submerged weight. Naturally, zero is imposed as the normal reaction when the cable is above the seabed.

If $\mathbf{V}_x$ is the vector velocity along the *x* axis, the friction force $\mathbf{F}^s$ due to the seabed acts along the same axis in the opposite direction of the node velocity (see also Liu and Bergdahl, 1997) and depends on the submerged weight of the lumped mass at the node.

To avoid numerical instabilities, this force is assumed to adopt a linear behaviour until the node speed is less than a small limit value, shown by $C_x$. Thus the expression shown in Figure 5 is as follows:

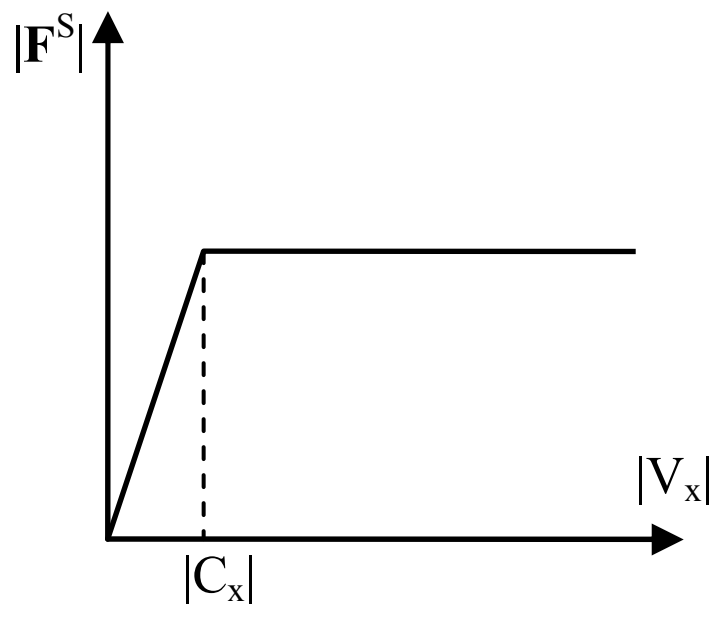

$$\begin{cases} F^s = -\mu \dfrac{V_x}{|C_x|} W & \text{if } V_x \le C_x \\ F^s = -\mu \dfrac{V_x}{|V_x|} W & \text{if } V_x > C_x \end{cases}$$

Figure 5. Modelling of the friction force at the seabed.

The slug flow regime is a highly complex phenomenon; therefore, when a complicated system such as a flexible marine riser subjected to a high degree of external force is analyzed, some simplifications have to be adopted to make the problem easier to study. Following the approach of many other authors interested in this type of problem, this study also uses a simple model of the slug flow regime.

In particular, methods similar to those of Patel and Seyed (1989) are used which describe the slug as a monophase liquid having a variable density characterized by a sinusoidal behaviour. Here, two simple models are presented: in the first, the effect of the pipe curvature on the slug regime will not be considered; the length of the Taylor bubbles and the liquid slugs the author aims to model are the same along the entire length of the pipe.

In both cases a monophase liquid is considered as a wave density travelling through the length of the riser, or rather as a train of liquid pulses that make an exertion on the inner pipe wall.

The biphasic flow is considered as a monophase fluid flow with variable density as a function of time and the riser's node location, as shown in Figure 6.

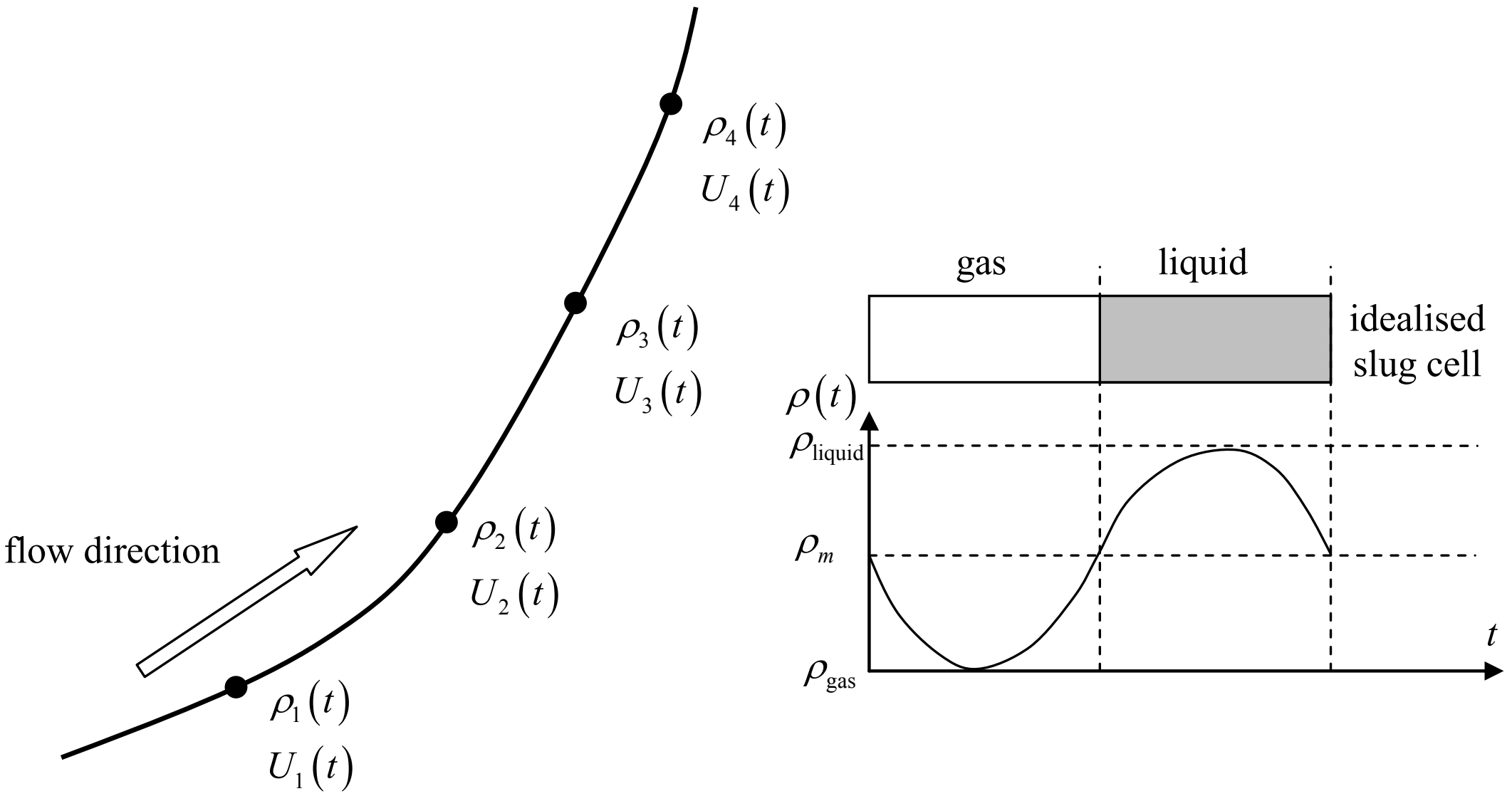

Figure 6. Simple scheme of the slug flow model.

At the section generic position $s$ the fluid density is modelled in the following way:

$$\rho(s,t) = \rho_m + \rho_f \cos(k\,s - \omega t) \tag{24}$$

where $\rho_m$ is the mean density of the fluid, $\rho_f$ its amplitude, $\omega$ the circular frequency of the slug flow regime, $t$ the time at which the density is evaluated, $s$ the curvilinear abscissa of the node and $k$ the slug wave number.

If $U_m$ denotes the mean velocity of the flow in a completely full pipe, the application of the continuity equation gives the fluid velocity at the generic section as:

$$U_i = \frac{\rho_m}{\rho_i} U_m \tag{25}$$

In the second case, a relationship between slug wavelength $L_s$ and the pipe inclination $\theta$ can be written.

The wave number $k$ and the pulsation $\omega$ in equation (24) are:

$$\begin{cases} k = \dfrac{2\pi}{L_s(\theta)} \\ \omega = \dfrac{2\pi}{T_s} = \dfrac{2\pi U_m}{L_s(\theta)} \end{cases} \tag{26}$$

If the maximum and minimum liquid slug wavelengths at the respective minimum $\theta_{min}$ and maximum $\theta_{max}$ pipe inclinations are imposed as (see also Fabre and Linè, 1992):

$$\begin{cases} L_{s_\max} = 30D_i \\ L_{s_\min} = 8D_i \end{cases} \tag{27}$$

then the following equation is obtained:

$$L_s(\theta) = \frac{\theta - \theta_{\max}}{\theta_{\min} - \theta_{\max}} \left( L_{s_\max} - L_{s_\min} \right) + L_{s_\min} \tag{28}$$

At the moment of $t$, the global dynamic equation of the motion applied to the control volume gives us (see Figure 7):

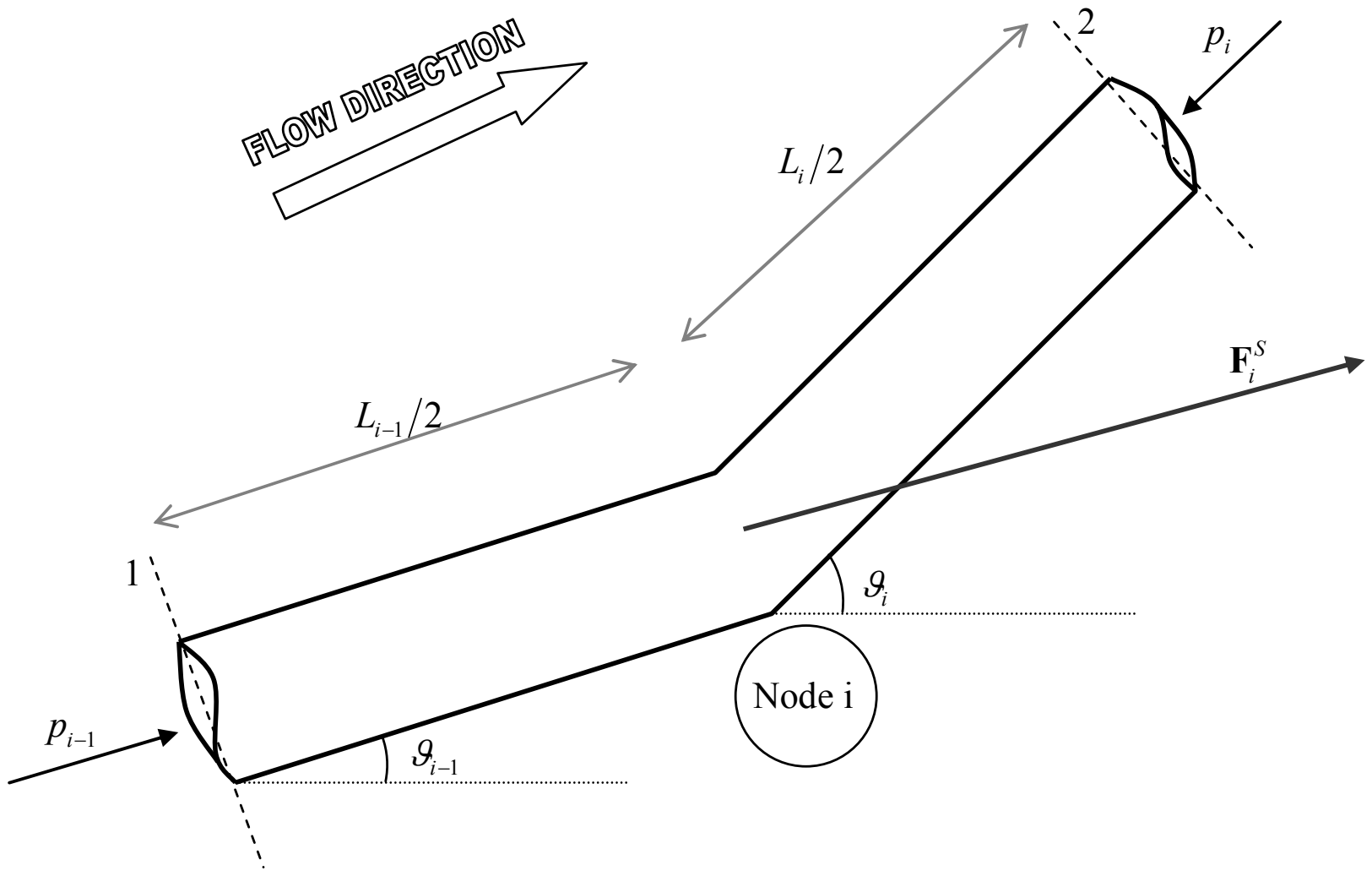

Figure 7. Control volume to which the global dynamic equation of equilibrium is applied.

$$\mathbf{\Pi} + \mathbf{G} + \mathbf{I} = \mathbf{M}_2 - \mathbf{M}_1 \Rightarrow \mathbf{\Pi}_1 + \mathbf{\Pi}_2 + \mathbf{F} + \mathbf{G} + \mathbf{I} = \mathbf{M}_2 - \mathbf{M}_1 \tag{29}$$

where $\mathbf{\Pi}_1$ and $\mathbf{\Pi}_2$ are the forces due to the pressure at sections *1* and *2*, while **F** is the force acting on the control volume through the lateral surface of the tube, **G** is the weight of the control volume and **I** represents the fluid local inertia force.

Thus the force due to the slug at the node is:

$$\mathbf{F}^S = -\mathbf{F} = \mathbf{M}_1 - \mathbf{M}_2 + \mathbf{\Pi}_1 + \mathbf{\Pi}_2 + \mathbf{G} + \mathbf{I} \tag{30}$$

It should be borne in mind that the pipe is located under seawater level and thus the external pressure due to the marine environment must also be evaluated in deriving the force acting on the pipe wall.

The local inertia force has the following expression:

$$\mathbf{I} = -\int_W \frac{\delta(\rho \mathbf{u})}{\delta t} dW = -\int_W \rho \frac{\delta \mathbf{u}}{\delta t} dW - \int_W \mathbf{u} \frac{\delta \rho}{\delta t} dW = -\int_W \rho \dot{\mathbf{u}}\, dW - \int_W \mathbf{u}\, \dot{\rho}\, dW \tag{31}$$

where the dot indicates the derivative with respect to time.

Equation (31) may be written by using the lumped mass approach applied to the control volume in Figure 7.

If $\mathbf{t}_{i-1}$ and $\mathbf{t}_i$ are the unit vector of elements *i-1* and *i* respectively, the unit vector associated with node *i* is expressed as:

$$\boldsymbol{\tau}_i = \frac{\mathbf{t}_j + \mathbf{t}_{j-1}}{\left|\mathbf{t}_j + \mathbf{t}_{j-1}\right|} \tag{32}$$

and the local inertia vector, lumped at node *i*, is estimated (in this study) as follows:

$$\mathbf{I}_i = -\left(\rho_i\, \dot{\mathbf{u}}_i\, \Delta W + \dot{\rho}_i\, \mathbf{u}_i\, \Delta W\right) = -\Delta W \left(\rho_i\, \dot{\mathbf{u}}_i + \dot{\rho}_i\, \mathbf{u}_i\right) \tag{33}$$

where:

$$\begin{cases} \rho_i = \rho_m + \rho_f \sin\left( \frac{2\pi}{L_s} s_i - \frac{2\pi}{T_s} t \right) \\ \dot{\rho}_i = -\frac{2\pi}{T_s} \rho_f \cos\left( \frac{2\pi}{L_s} s_i - \frac{2\pi}{T_s} t \right) \end{cases} \tag{34}$$

By considering equation (25) it is:

$$\dot{u}_i = \rho_m u_m \frac{d}{dt} \frac{1}{\rho_i} = \rho_m u_m \frac{d}{dt} \rho_i^{-1} = \rho_m u_m \left( -\rho_i^{-2} \left( -\frac{2\pi}{T_s} \rho_f \cos\left( \frac{2\pi}{L_s} s_i - \frac{2\pi}{T_s} t \right) \right) \right) =$$

$$= \rho_m u_m \frac{2\pi}{T_s} \rho_f \frac{\cos\left( \frac{2\pi}{L_s} s_i - \frac{2\pi}{T_s} t \right)}{\left( \rho_m + \rho_f \sin\left( \frac{2\pi}{L_s} s_i - \frac{2\pi}{T_s} t \right) \right)^2} \tag{35}$$

thus:

$$\rho_i \dot{u}_i = \rho_m u_m \frac{2\pi}{T_s} \rho_f \frac{\cos\left( \frac{2\pi}{L_s} s_i - \frac{2\pi}{T_s} t \right)}{\rho_m + \rho_f \sin\left( \frac{2\pi}{L_s} s_i - \frac{2\pi}{T_s} t \right)} \tag{36}$$

But it is also:

$$\dot{\rho}_i u_i = -\frac{2\pi}{T_s} \rho_f \cos\left( \frac{2\pi}{L_s} s_i - \frac{2\pi}{T_s} t \right) \frac{\rho_m u_m}{\rho_m + \rho_f \sin\left( \frac{2\pi}{L_s} s_i - \frac{2\pi}{T_s} t \right)} \tag{37}$$

thus:

$$\dot{\rho}_i u_i = -\rho_m u_m \frac{2\pi}{T_s} \rho_f \frac{\cos\left( \frac{2\pi}{L_s} s_i - \frac{2\pi}{T_s} t \right)}{\rho_m + \rho_f \sin\left( \frac{2\pi}{L_s} s_i - \frac{2\pi}{T_s} t \right)} \tag{38}$$

Subsequently, by comparing equations (36) and (38) it was found that:

$$\rho_i \dot{u}_i = -\dot{\rho}_i u_i \qquad (39)$$

Decomposing the acceleration along the tangential and normal directions and referring to result (39), the local inertia may be rewritten as:

$$\mathbf{I}_i = -\Delta W \left( \rho_i \dot{u}_i \boldsymbol{\tau}_i + R u_i \frac{d}{dt}\left(\frac{1}{R}\right) \boldsymbol{\tau}_i + \rho_i \frac{u_i^2}{R} \mathbf{n}_i + \dot{\rho}_i u_i \boldsymbol{\tau}_i \right) = -\Delta W \left( -\frac{u_i}{R} \frac{dR}{dt} \boldsymbol{\tau}_i + \rho_i \frac{u_i^2}{R} \mathbf{n}_i \right)$$

(40)

in which the curvature variation is evaluated as its discrete variation between two temporal steps.

Following the approach of Patel and Seyed (1989), the force expressed by equation (30) must be adjusted by considering the hydrostatic pressure force due to seawater acting on the outer pipe wall. Therefore, the final expression of the dynamic equation is:

$$\begin{cases} f_x = p_1 A_i \cos\vartheta_1 - p_{1_ext} A_e \cos\vartheta_1 - p_2 A_i \cos\vartheta_2 + p_{2_ext} A_e \cos\vartheta_2 + \\ +\rho_1 U_1^2 A_i \cos\vartheta_1 - \rho_2 U_2^2 A_i \cos\vartheta_2 \\ f_y = p_1 A_i \sin\vartheta_1 - p_{1_ext} A_e \sin\vartheta_1 - p_2 A_i \sin\vartheta_2 + p_{2_ext} A_e \sin\vartheta_2 + \\ +\rho_1 U_1^2 A_i \sin\vartheta_1 - \rho_2 U_2^2 A_i \sin\vartheta_2 - G \end{cases} \qquad (41)$$

where $p_{1_ext}$ and $p_{2_ext}$ represent the external seawater pressure at sections *1* and *2* respectively.

The total inner pressure $p_i$ at point *i* may be represented as the sum of the stationary pressure $p_{is}$ as a result of the steady component of the flow and the variable pressure $\Delta p_i$ resulting from density fluctuations:

$$p_i = p_{is} + \Delta p_i \qquad (42)$$

The steady pressure is evaluated by considering the mean values of the velocity and the equivalent fluid density. If the Bernoulli equation, in which the energy loss term is included, is applied between two generic points *1* and *2*, and the flow passes from point *1* to point *2*, then it is:

$$p_{2s} = p_{1s} + \left(y_1 - y_2\right)\rho_m\, g - J\, L_{12}\, \rho_m\, g \tag{43}$$

where $J$ is the energy loss per unit length, $y$ the depth of the point, $L_{12}$ the distance between the points and $\rho_m$ the fluid mean density. By supposing an extremely turbulent flow, the Darcy-Weisbach equation, in which Darcy's friction factor $\lambda$ appears, may be used:

$$J = \lambda \frac{U^2}{2\, g\, D_i} \tag{44}$$

By assuming all of the quantities as constant (internal diameter, relative roughness, mean velocity), $J$ will be constant in each equation and it will depend only on $\lambda$.
If a very high Reynolds number is supposed (hydraulically rough pipe), the Prandtl-von Karman equation may be used:

$$\frac{1}{\sqrt{\lambda}} = -2\log\left(\frac{1}{3.71}\frac{\varepsilon}{D_i}\right) \tag{45}$$

where $\varepsilon$ is the equivalent roughness of the pipe.
The fluctuations of the pressure at each section may be evaluated by considering the weight of the fluid above the node considered at each time step.
The variable pressure at the section is:

$$\Delta p_i = \int_{s_i}^{s_k} g\, \rho_f\, A \sin\left(k\, s - \omega t\right) = \frac{1}{k} g\, \rho_f\, A\left[\cos\left(k\, s_i - \omega t\right) - \cos\left(k\, s_k - \omega t\right)\right] \tag{46}$$

When the inner flow into the pipe is considered, the weight of the fluid must be considered. This weight is the sum of the steady weight due to the mean density and the fluctuating weight due to the fluctuating density component. The submerged weight is still considered in terms of the external volume occupied by the pipe.
The steady weight is:

$$G_s = \rho_m \, g \, A_i \, L_{12} \tag{47}$$

where $L_{12}$ is the element length.
The variable weight between the generic section *i* and the following section *i+1* of the riser is:

$$\Delta G = \int_{s_i}^{s_{i+1}} \rho_f \, g \, A_i \sin\left(k\,s - \omega t\right) ds = \frac{1}{k} \rho_f \, g \, A_i \left[\cos\left(k\,s_i - \omega t\right) - \cos\left(k\,s_{i+1} - \omega t\right)\right] \tag{48}$$

It must be highlighted that in the previous description, a precise analysis must take into account the relative velocity between the inner fluid and the structural velocity. In this paper this aspect was not considered, thus more simple analysis is used in which the slug velocity refers to a fixed pipe. Further improvements of the codes will also allow the author to take this effect into account.
Patel and Seyed (1989) highlight that the mean period of a slug flow is about 4s which is low compared to a typical wave period (for example 10s). The implication is that fatigue breaks resulting from higher slug frequencies are more important than those from wave action.
Furthermore, wave loads, which have low frequencies in open sea, primarily affect pipe flexure, its bending moment distribution. Slug flow regime, on the other hand, acts on higher frequencies and has a dominant tensile effect; it acts primarily upon the axial effective tension distribution. It is for this reason that the simulations carried out in this study used slug flow with a relatively high frequency.

## DESCRIPTION OF THE SIMULATIONS

In the present paper a series of simulations were carried out in order to appreciate differences in the pipe behaviour with the aforementioned two slug flow models.

In order to highlight only these aspects, a number of parameters were considered constant in each simulation, which had a real time of 400s, and the output was recorded every 0.1s. Table 1 shows the main quantities and the values which characterised the input parameters.

Table 1. Main constant quantities used for the simulations

| | | |
|---|---|---|
| **Cable length** | **[m]** | **1400** |
| **Number of elements** | **[-]** | **80** |
| **Horizontal span** | **[m]** | **1200** |
| **Vertical span** | **[m]** | **550** |
| **Axial rigidity EA** | **[N]** | **$5 \cdot 10^8$** |
| **Moment of inertia EJ** | **[$kN \cdot m^2$]** | **120.8** |
| **Inner diameter** | **[m]** | **0.305** |
| **Outer diameter** | **[m]** | **0.396** |
| **Critical damping factor** | **[-]** | **0.2** |
| **mass per unit length** | **[kg/m]** | **165** |
| **Drag coefficient** | **[-]** | **1** |
| **Normal added mass coefficient** | **[-]** | **1** |
| **Seawater density** | **[$kg/m^3$]** | **1024** |

The global equation of the motion of the cable was calculated starting from single equations which refer to the various elements of the cable and the code was computed by using the algorithm ODE113 as implemented in Matlab, with the proper boundary and initial conditions.

The code was validated in the absence of a seabed and slug force as presented in a previous paper by the authors (Pollio et al., 2006).

The main flaw of the implemented algorithm was the slowness of the simulation in the time domain, but it was inherently stable over a number of conditions. One reason for using it is the fact that many authors prefer to use the classic Newmark method in order to simulate such a system. This study may therefore be useful as a means of comparison with existing or future research.

Since the code does not take into account the variable submerged length of the upper elements when they are subjected to wave motion, the top node of the riser was set at 5m under the mean seawater surface. As mentioned before, the system underwent a simple sinusoidal wave motion as described by Airy's theory. In all cases, the wave had a height of 6m and a time period of 10s.

The seabed was flat and its depth corresponded to the depth of the bottom node in order to keep the riser lying on it at its lower parts. A simple sketch of the system is shown in Figure 8.

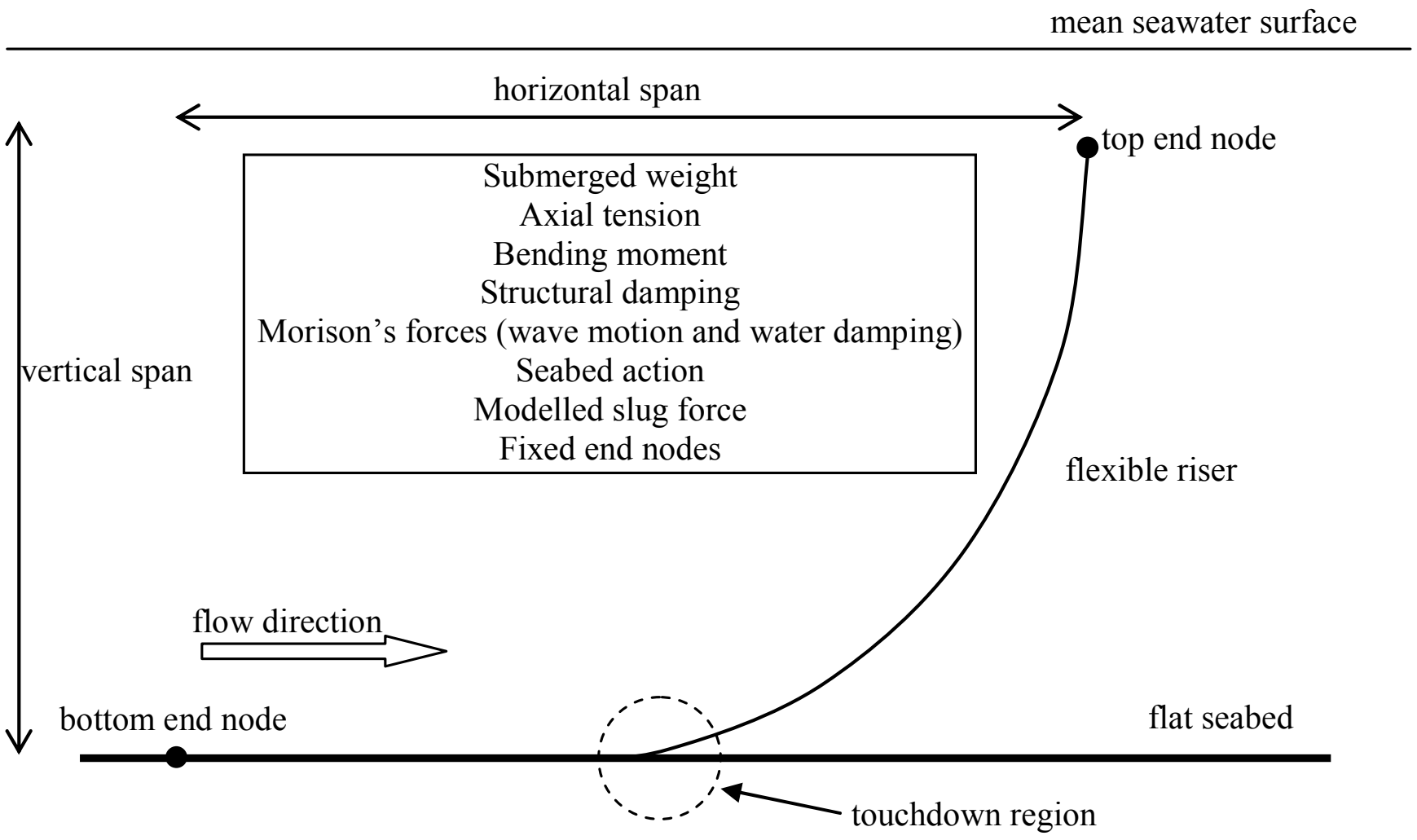

Figure 8. A simple sketch of the riser configuration with the presence of the seabed and all the acting forces.

The riser initially had an elastic catenary configuration, which is very close to a static equilibrium configuration (at least for the configuration in the absence of a seabed) even though at the initial temporal steps of the simulations, a transient period developed and was characterized by clearly pronounced peaks in the time domain.

The movements at the end nodes were not considered (or rather, platform or ship movements were not modelled here), therefore the configuration is that of a hanging riser between two fixed points.

An important topic in research on risers is that of pipe behaviour in correspondence to the touchdown point, the point from which the riser detaches itself from the seabed. In this work a detailed analysis of this aspect is not considered, but it is clear that over the points in this region the bending moment undergoes an abrupt variation, which implies a very high exertion on the material of the riser.

Table 2 shows the main parameters of the two sets of simulations for both the slug flow models presented here. It can be seen that all the parameters were used for simulations both with and without the seabed. In the first case, the free hanging riser was subjected to all forces except seabed reaction. In the case of constant slug density along the riser, only the slug wavelength varies and a constant mean velocity of 6m/s was used as a typical value proposed by Patel and Seyed (1989). In the case of the slug density variable with pipe inclination, the slug wavelength assumed the expression presented in the previous section and also in this case the mean travelling velocity of the fluid was set at 6m/s.

Since the inner diameter is 0.305m, following equation (27), the maximum and minimum lengths of the slug cell were 10.5m and 2.44m respectively with a difference of 8.06m. Thus the 8 simulations concerning the presence of slugs with constant wavelengths were characterized by slug wavelengths from the value of 2.44m to 10.5m, as seen in Table 2. In particular, the tests in which the seabed is absent are given the tag *Ti* while *TBi* is the tag for tests characterized by the presence of a seabed (*i* goes from 1 to 7). The density of the equivalent liquid of the biphasic slug regime was based on a liquid density of 900kg/m3 and a gaseous density of 500kg/m3 (this high value was chosen taking into account that, generally, the gaseous phase inside the pipe is under pressure): the equivalent liquid had a sinusoidal density with values ranging between these. In order to avoid a scenario where the inner flow pressure was less than the outer seawater pressure, at the bottom end node the starting inner fluid pressure was set as the sum between the maximum outer pressure along the pipe, obtained by supposing an hydrostatic pressure distribution, and a reference pressure of $10^6$Pa.

Table 2. Tests carried out.

| **Simulation name** | **Slug wavelength [m]** | **Slug frequency [Hz]** | **Slug period [s]** |
|---|---|---|---|
| **T1, TB1** | **2.44** | **2.46** | **0.41** |
| **T2,TB2** | **5.12** | **1.17** | **0.85** |
| **T3,TB3** | **7.80** | **0.77** | **1.30** |
| **T4,TB4** | **10.5** | **0.57** | **1.75** |
| **T5,TB5** | **Slug flow with frequency from 0.57 Hz to 2.46 Hz depending on pipe inclination** | | |
| **T6,TB6** | **Fluid with constant density** | | |
| **T7,TB7** | **No inner fluid** | | |

## RESULTS

The current objective is to compare riser behaviour when it undergoes a slug phenomenon either with constant slug wavelength value (that is, with a density independent of riser inclination) or varying with the pipe inclination, in the presence or absence of the seabed. In the latter case, the author aims to investigate the influence of the sea bottom in the two slug models. The two geometrical configurations are sketched in Figure 9.

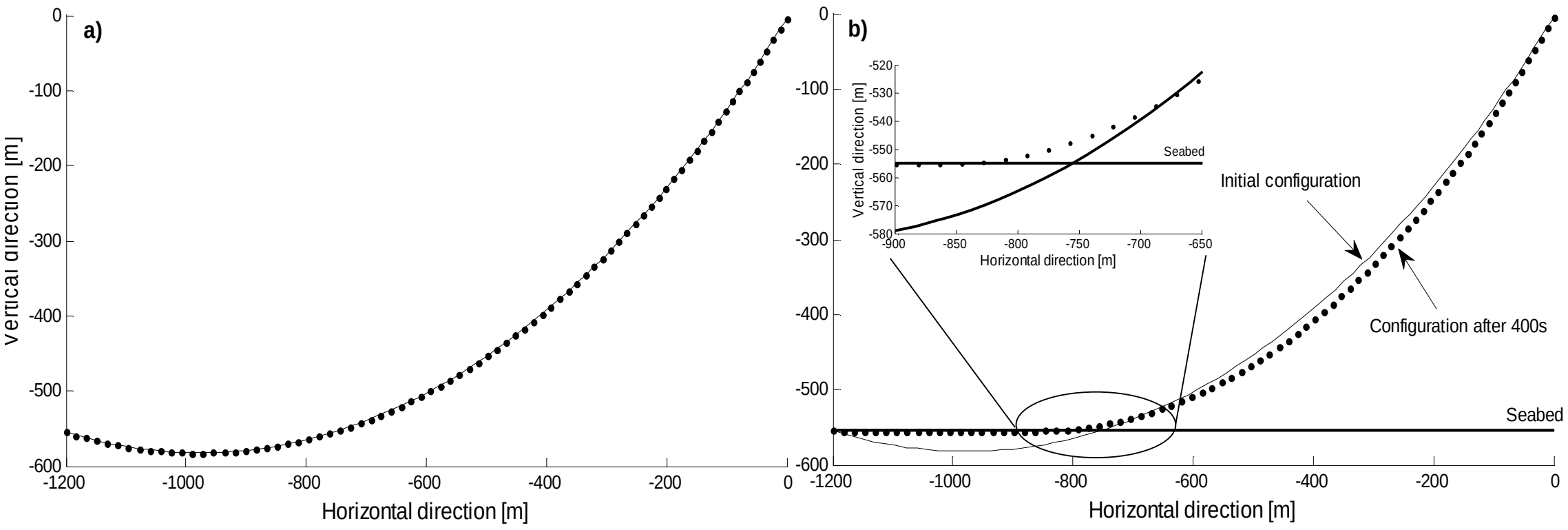

Figure 9. Examples of the geometric configurations of the riser in absence of bed (T1, a) and in presence of bed (TB5, b). (continuous line: initial riser geometry; dotted line: configuration after 400s).

Figure 10 shows the top effective tension temporal behaviour for all the Ti tests. It is possible to note that the presence of a slug flow with density as a function of the inclination (T5) affects the axial tensile force in the same way as the effects of the slug flow with constant slug wavelengths (T1-T4) in the same range of frequencies. Moreover, an inner flow with constant density (T6) gives practically the same mean top effective tensions. In the absence of inner fluid, the tension values are very low in comparison with their former ones (T7). Figure 10-b shows a close-up over the last 50s of the simulations in which it is possible to see the configuration with a constant fluid density (T6) presenting a top tension with a main period of 10s, the same as the wave acting upon the pipe. In the other cases (T1-T4 and T5), the riser is subjected to more complex tension variations due to the intrinsic frequency of the liquid slugs passing through the riser. In all cases the tension behaviour seems to have a good general regularity.

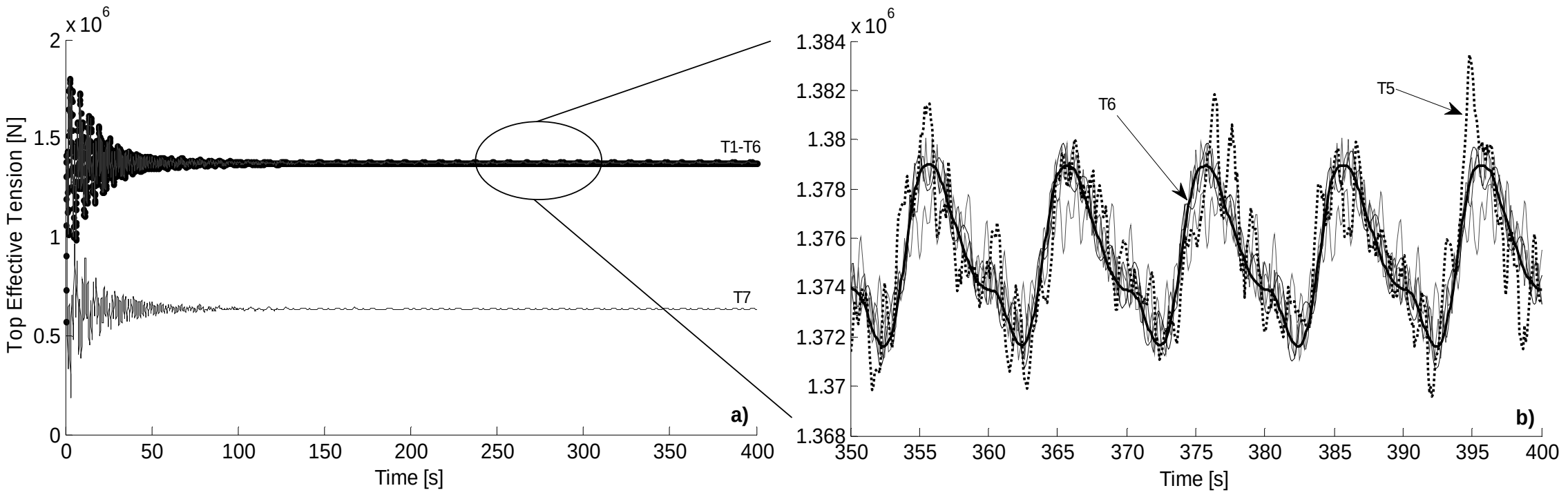

Figure 10. Top effective tension temporal behaviour for all the Ti tests with a close-up of the last 50s (continuous line, T6; dotted line, T5).

Similar results are obtained for the bottom effective tension, which is the tension evaluated at the bottom end node, as Figure 11 shows. Also in this case, test T5 gives the same tension as the tests T1, T2, T3, T4 and T6. A close-up (Figure 11-b) highlights this phenomenon clearly. Figure 11 shows that T5 results are much more disturbed than the others.

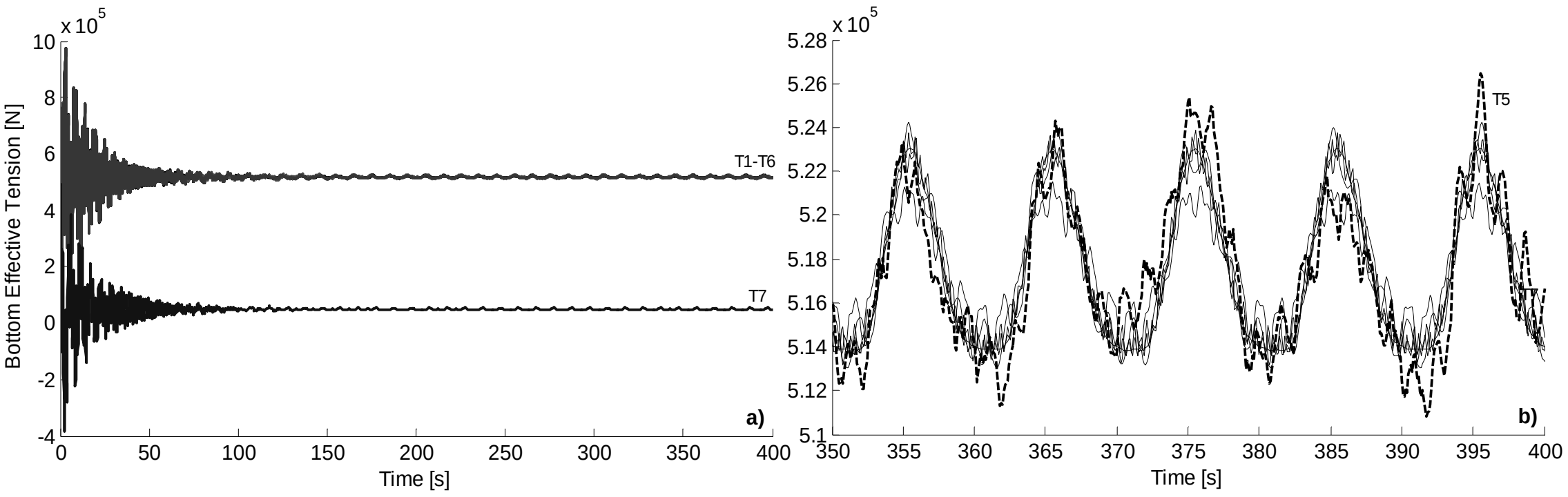

Figure 11. Bottom effective tension temporal behavior for all the Ti tests with a close-up of the last 50s (continuous line, (T1-T4,T6); dotted line, (T5)).

The greater distortion of the axial tension due to the presence of a slug flow with a varying frequency along the entire pipe length (T5) is shown in Figure 12 where its irregular trend is clearly visible (b) in comparison to the more regular behaviour of all the other tests (a). This phenomenon is almost certainly due to the continuous variable pulling that the liquid slugs exert along the pipe length at the different locations.

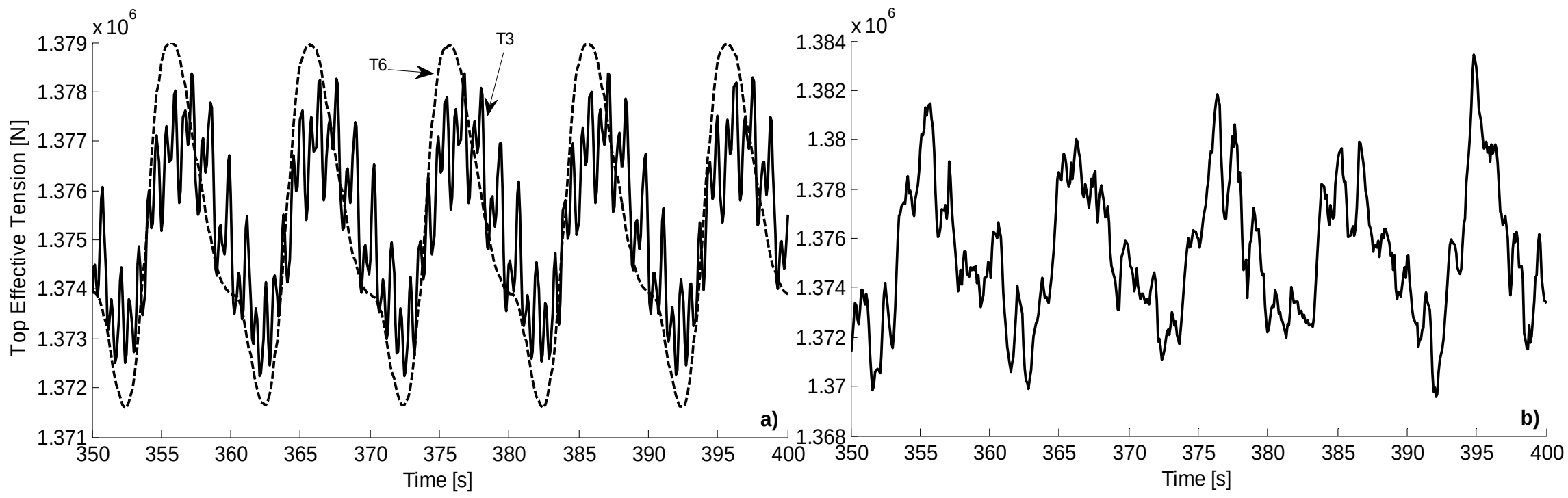

Figure 12. Top tension behavior in the presence of slugs with varying density (T5, a) and a comparison with tests T1, and T6 (b).

The frequency analysis carried out for all the tests always shows peaks in correspondence with the forcing frequencies acting upon the system. As an example, Figure 13 highlights the peaks at 0.1Hz for tests T6 and T2 due to the frequency of the wave motion (present in all tests). This frequency can also be seen in the time domain by observing the general trend of the tensions (see Figure 12-b as an example). Furthermore, for test T2, the peak is 1.17Hz, which is the frequency of the liquid slugs flowing into the pipe.

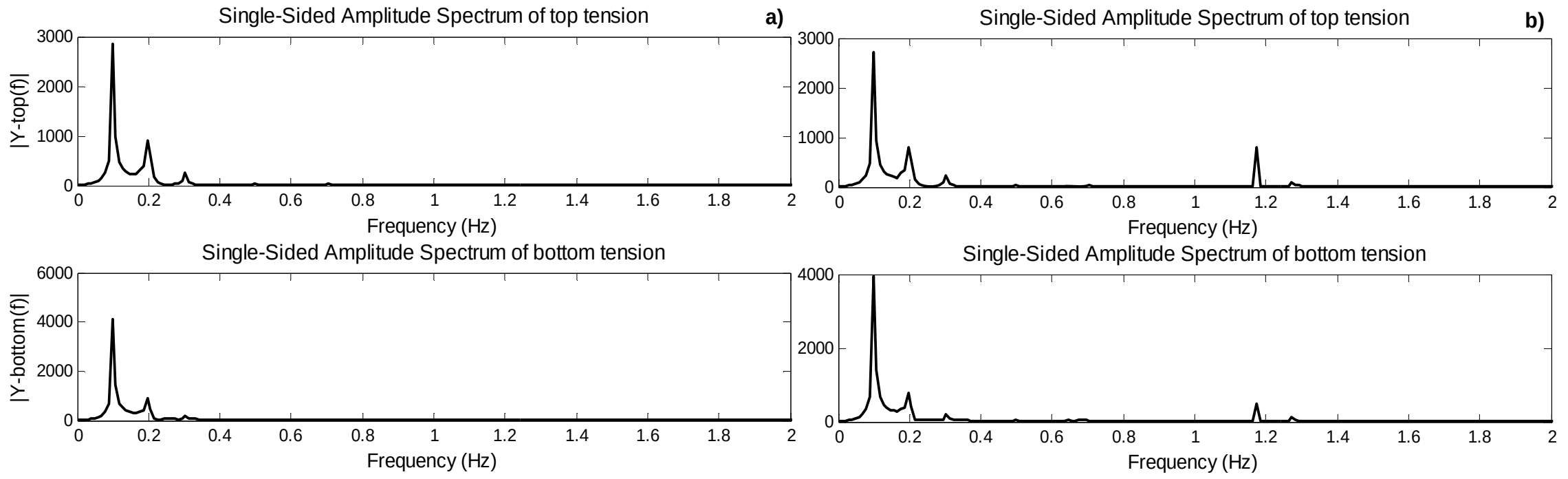

Figure 13. Frequency analysis at the top and bottom end node for test T6 (a) and test T3 (b).

It is interesting to note from Figure 14-a and Figure 14-b that the mean effective tension and the mean bending moment distributions obtained in test T5 have a general behaviour practically identical to that obtained from the other tests. A similar conclusion is also obtained in the presence of the seabed, as will be seen.

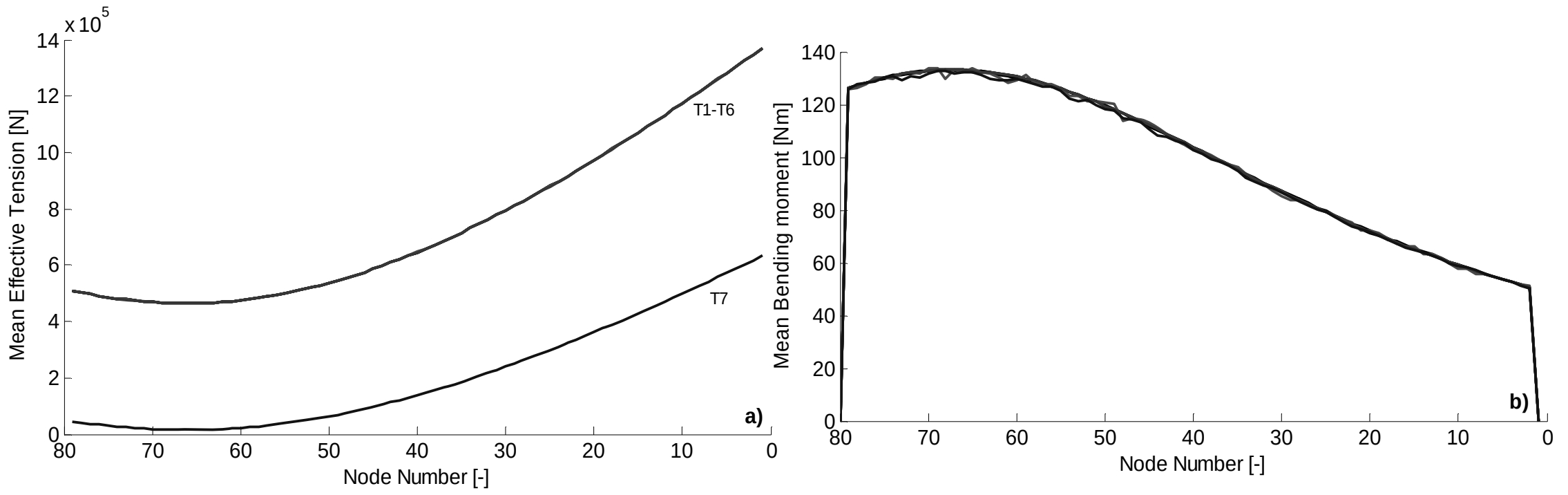

Figure 14. Mean effective tension (a) and mean bending distributions (b) along the riser for the Ti tests.

The differences between the maximum and the minimum effective tension and bending moment obtained along the riser are reported in Figure 15. Test T5 presents the higher variations of tension between its minimum and maximum values. This does not happen with the same magnitude in the other tests (Figure 15-a), while test T3 has the highest variation of the bending moment along the riser. Such peculiar behaviour may be due to the possible resonant effect at the frequency of almost 0.77Hz that mainly affects the moment values, even though the study of the natural frequency of the riser is not presented here.

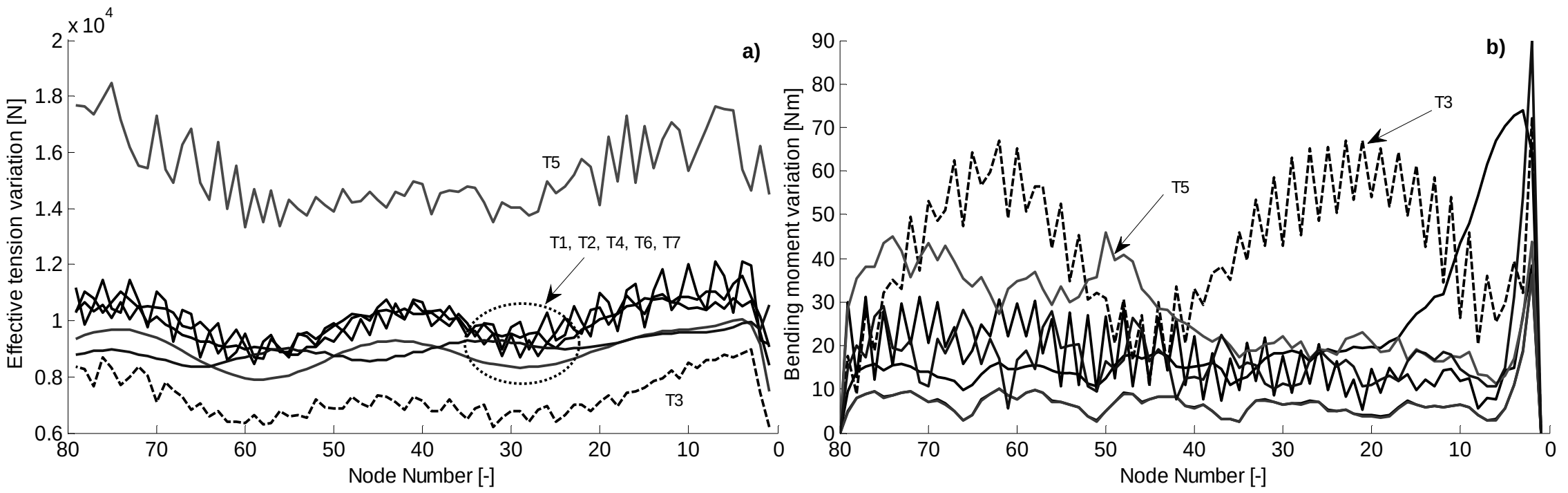

Figure 15. Variations of the effective tension (a) and the bending moment (b) along the riser for the Ti tests.

The results obtained from the simulations with seabed presence show that the axial tension is less than that obtained in the absence of the bed at both the pipe extremities. In particular Figure 16 highlights that, also in this case, the top effective tension for test TB7 (no presence of inner fluid, and pipe subjected only to wave motion) is less than that of the others. In this case, a close-up of Figure 16 shows that test TB3 (slug frequency of 0.77Hz) has a greater effect on the values of the maximum and minimum effective tension. Figure 17 shows that at the bottom end node, the effective tension values also are practically the same for all the tests, except test T7, in which there is no inner flow.

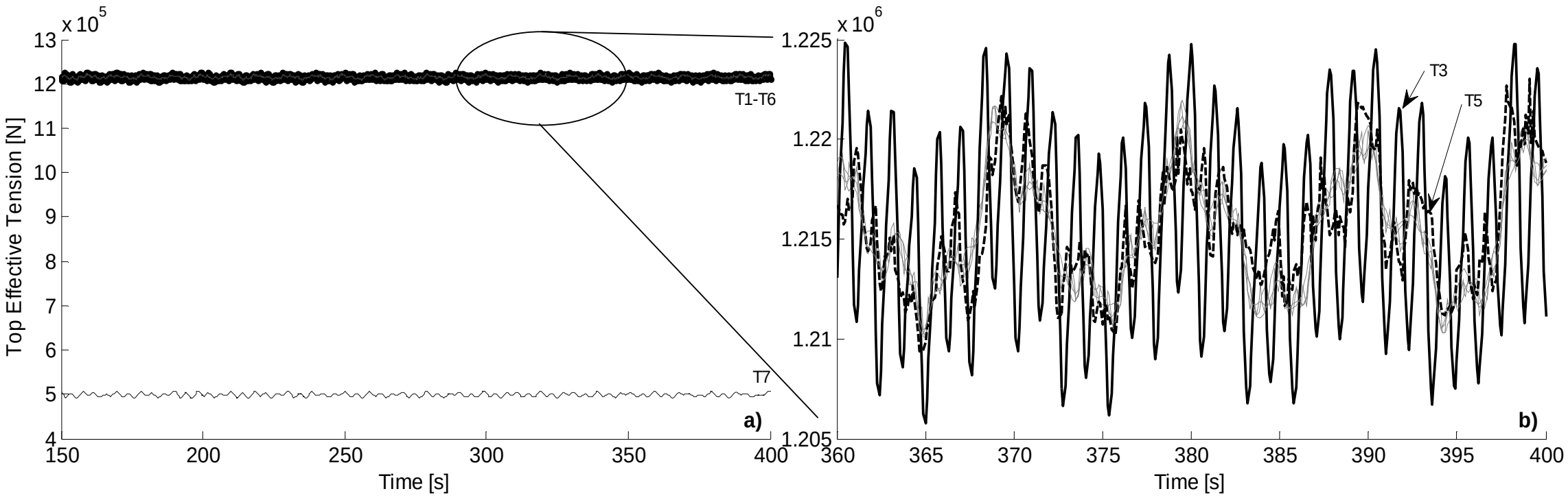

Figure 16. Top effective tension temporal behavior for all the TBi tests with a close-up of the last 50s (continuous line, (T3); dotted line, (T5)).

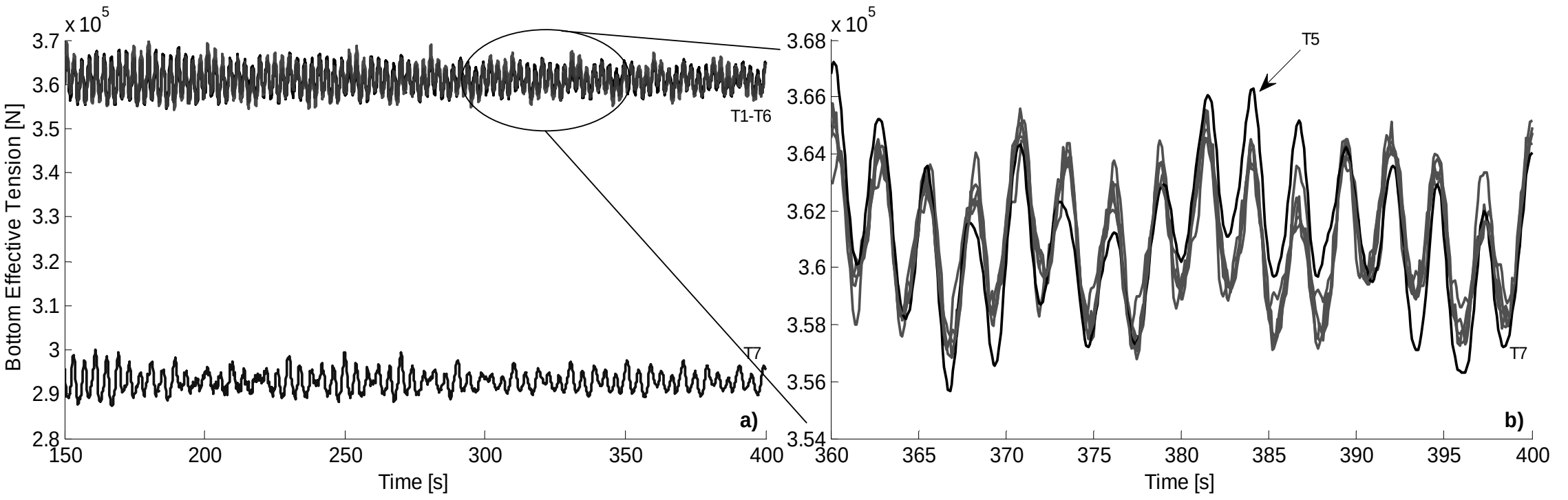

Figure 17. Bottom effective tension temporal behavior for all the TBi tests with a close-up of the last 50s.

Figure 18 presents the mean effective tension and moment distribution along the riser lying on the seabed. Figure 18-a gives similar results to those obtained in the absence of the seabed, while Figure 18-b highlights the presence of the touchdown region at which the bending moment values increase suddenly due to the riser detachment from the seabed.

In this case, all the tests carried out show a similar behaviour of the moment which is very slight for the points lying on the seabed and increases to relatively high values that may affect the resistance of the riser when it is subjected to flexing exertions.

In the presence of the seabed, too, the variations for the effective tension and the moments along the pipe are plotted (Figure 19-a and Figure 19-b respectively). In this case, the differences are lower than those obtained for the Ti tests (see Figure 15); this may well be an important aspect that must be considered in designing this kind of pipe when a seabed-cable interaction is present. Furthermore, the slug flow with variable density affects the pipe with higher variation of the axial tension between its maximum and minimum values (except in the case of test TB3), and this fact appears to be extremely important in the fatigue analysis of marine risers. It is interesting to note that also in this case, test TB3 presents the greater variation of the axial tension and the bending moment, and this may be more possible evidence of a resonant frequency of the system which is very close to 0.77Hz.

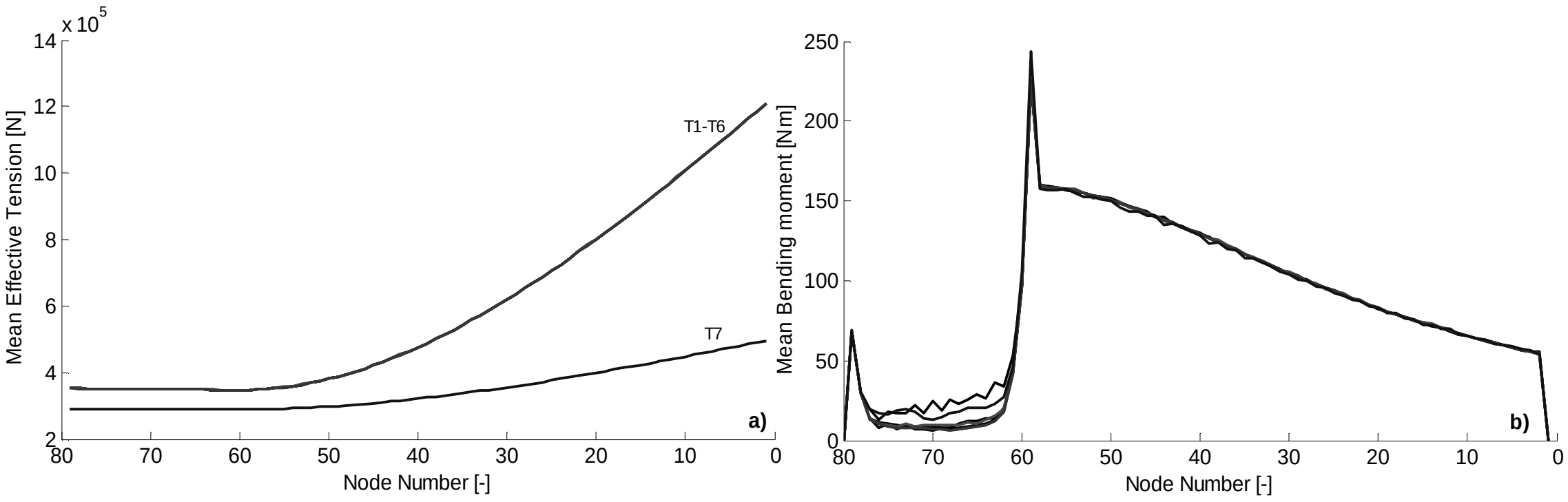

Figure 18. Mean effective tension (a) and mean bending distributions (b) along the riser for the TBi tests.

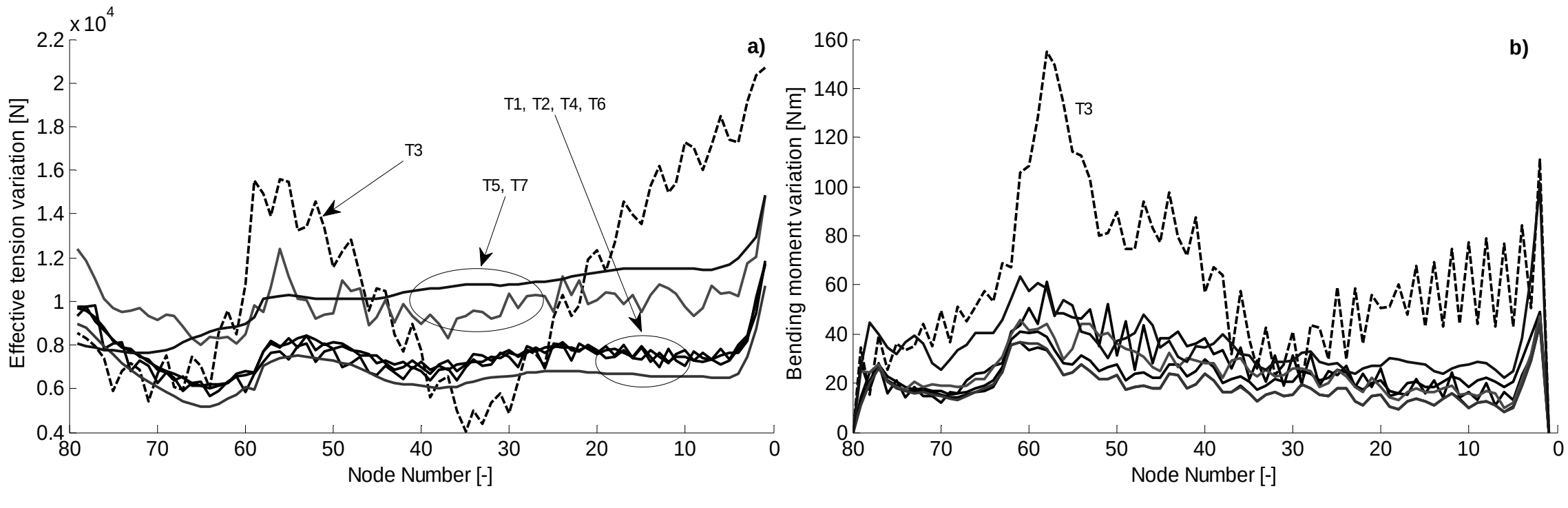

Figure 19. Variations of the effective tension (a) and the bending moment (b) along the riser for the TBi tests.

## CONCLUSIONS

In this study the authors present a comparison between two simple models of slug flow through a very long flexible marine riser, both in the presence and absence of a flat seabed.

The system was modelled by considering the so-called lumped mass approach in a bidimensional reference system lying in the same plane of the studied pipe. Specifically, the modelled forces acting upon the pipe in this study are the submerged weight of the riser, the Morison's forces in the presence of regular monochromatic waves, the slug flow forces and the seabed-riser interaction. They allowed for the study of the pipe inner stresses in terms of axial tensile force and bending moment. A global non-linear equation of motion for the entire system, together with proper boundary and initial conditions, was created starting from the equations of the motion of the single elements compounding the riser. The slug flow was modelled as a travelling density wave with a wavelength independent of the function on the pipe inclination.

A series of time domain simulations were carried out using the ODE 113 algorithm implemented in MATLAB.

Some interesting results obtained from the simulations can be summarised as follows:

- In the presence of a slug flow with varying wavelength, the axial tensile force at the top and bottom end nodes of the riser shows practically the same values as those obtained from scenarios where a steady flow or slug flow with a constant frequency travel along the pipe.
- a slug flow with variable frequency allows for irregular inner stress behaviour over time, while the tension and moment variations seems to be more regular when a steady flow and slug flow with constant frequency occur.
- In the presence of the seabed there are fewer differences between the maximum and minimum values of the axial tension along the riser with respect to those obtained in the absence of the seabed for all the tests. The opposite phenomenon happens to the variation between the maximum and the minimum values of the bending moments, as a comparison between Figure 15 and Figure 19 shows. Furthermore, the presence of a slug flow with variable frequency either with seabed-riser interaction or without it has a greater effect on these differences and the riser itself undergoes higher stress with a greater probability that it will reach failure point.

## ACKNOWLEDGEMENTS

The authors would like to thank prof. Robin Langley of the University of Cambridge (UK), Department of Engineering, Dynamics and Vibration Group, and prof. Giuseppe Marano, Technical University of Bari, Department of Environmental Engineering and Sustainable Development, for their help in the development of the present code.